\documentclass[10pt,epsf]{article}

\begin{document}

\title{Collisional relaxation in a fermionic gas}

\author{Gabriele Ferrari}
{\it Laboratoire Kastler Brossel, D\'epartement de Physique de l'Ecole Normale Sup\'erieure,\\ 24 rue Lhomond, 75231
Paris CEDEX 05, France} \maketitle
\begin{abstract}
We propose a method to study the degeneracy of a trapped atomic gas of fermions
through the relaxation of the motion of a test particle. In the degenerate
regime, and for an energy of the test particle well below the Fermi energy, we
show that the Fermi-Dirac statistics is responsible for a strong decrease of
the relaxation rate. This method can be used to directly measure the
temperature of the fermionic gas.
\end{abstract}

\vspace{5mm} PACS numbers: 05.30.Fk, 03.75.Fi, 32.80.Pj

New techniques for cooling atomic gases, based either on laser manipulation or
evaporative cooling, have led to spectacular progresses in the realization of
degenerate atomic samples. The most striking example is the Bose-Einstein
condensation of alkali vapors \cite{anderson,davis,bradley} and more recently,
of an atomic hydrogen gas \cite{Kleppner}. These techniques can also be applied
to fermionic samples, in which spectacular phenomena such as Cooper pairing of
atoms \cite{Stoof1}, or inhibition of the spontaneous emission of an excited
atom \cite{Helmerson} have been predicted. In this article, we present a simple
and powerful tool to study the degeneracy parameters of such a fermionic
sample, which we consider ideal for simplicity.

Numerous possible ways to analyze an atomic Fermi gas in the degenerate regime have been recently studied from an
theoretical point of view. Several authors focus on the interaction of the atomic cloud with light: modifications of
the refraction index or of the absorption coefficient of the gas \cite{Morice,Javanainen}, reduction of the spontaneous
emission rate of an excited atom inside the atomic cloud and angular dependence of the radiation pattern
\cite{busch,demarco}. Spectacular effects are expected when the Fermi momentum $p_F$ of the cloud is larger than the
photon momentum. Unfortunately in this regime, a quantitative analysis of the interaction of the gas with resonant
light is difficult. Indeed it corresponds to $n\lambda^3 >1$, where $n$ is the spatial density of the gas and $\lambda$
is the wavelength of the light. Light propagation in the medium is then strongly affected by multiple scattering
effects, which are difficult to handle.

Other propose diagnostics of Fermi degeneracy of a trapped gas involve the
study of its spatial distribution, either for a pure fermionic sample
\cite{silvera,butts,schneider,burnett}, or for a mixture with a Bose-Einstein
condensate \cite{moelmer}. These methods are well adapted to study the region
where the temperature $T$ is around the Fermi temperature $T_F$, but
 become less sensitive to temperature in the strongly
degenerate regime where $T\ll T_F$.
Also, as pointed out in \cite{demarco}, one
can take advantage of inelastic
processes inside the gas to collect
 information about its degeneracy
parameter.

In the present paper we investigate the dynamics of a probe particle (P) in the
same trapping potential as the Fermi gas. We assume that we can prepare P on an
arbitrary orbit of the trapping potential. An experimental procedure for such a
preparation is outlined in the last part of this paper. Neglecting any
inelastic process, the only way to change the trajectory of P is an interaction
with the Fermi gas. We show that the collisional dynamics of P gives access to
the statistical properties of the Fermi gas, in particular its temperature $T$.

The trapping potential for the test particle $U({\bf r})$ is assumed to be
harmonic and isotropic, with a frequency $\omega/2\pi$. For simplicity we
neglect the modification to the potential due to the mean field interaction
between P and the Fermi gas. We suppose also that the same trapping potential
acts on the Fermi gas. We calculate the dynamics of P in the Fermi sea by means
of the Boltzmann transport equation \cite{huang} in the local density
approximation:
\begin{equation}
\frac{\partial w}{\partial t}+ {\bf v}\cdot{\bf \nabla}_{\bf r} w
 + {\mathbf \nabla}_{\bf r}U \cdot{\bf \nabla}_{\bf p}
w = \left.\frac{d w }{d t}\right|_{\rm coll.}
\end{equation}
where $w({\bf r},{\bf p},t)$ is the phase space density of p at time $t$. We
put ${\bf v}={\bf p}/M$, where $M$ is the mass of P. With a notation similar to
\cite{huang} the collisional contribution to the quantum Boltzmann transport
equation for P is:
\begin{equation}
\left.\frac{d w }{d t}\right|_{\rm coll.}({\bf r},{\bf p},t)=
-\frac{\sigma}{4\pi h^3}\int d^3p_f \; d^2 \Omega\;
\left(w f(1-f')-w' f' (1-f)\right)\; |{\bf v}-{\bf v}_f|
\label{Bolt}
\end{equation}
where $\sigma$ is the collisional cross section for interactions between P and
fermions. Collisions occur in the low energy range where $s$-wave scattering is
dominant hence the cross-section is isotropic and independent of energy.
  The first part of the collisional integral corresponds to the decay
of the phase space density in ${\bf r},{\bf p} $ due to a collision between P and a fermion with momentum ${\bf p}_f$:
${\bf p} + {\bf p}_f \rightarrow {\bf p'} + {\bf p'}_f$. The final relative velocity is pointing in the direction given
by $\Omega$. The second part of the integral describes the reverse process, and we put ${\bf v}_f={\bf p}_f/m$, where
$m$ is the fermionic mass. We assume that a single probe particle is present, or that the gas of probe particles is
sufficiently dilute to be treated as an ideal Boltzmann gas. In the latter case we suppose that the number of test
particles is sufficiently low that the relaxation does not significantly perturb the distribution of the Fermi gas. We
use the abbreviations $w'=w({\bf r},{\bf p'},t)$, $f=f_T({\bf r},{\bf p}_f)$ and $f'=f_T({\bf r},{\bf p'}_f)$. The
quantity $f_T({\bf r},{\bf p})$ represents the steady-state phase space distribution of the Fermi gas in the local
density approximation:
\begin{equation}
f_T({\bf r},{\bf p})=\frac{1}{1+\exp\left(
\left( p^2/2m+U(r)-\mu\right)/(k_B T)\right)}
\end{equation}
where $\mu$ is the chemical potential. The Pauli exclusion principle is
represented in the collisional integral by the factors $1-f'$ and $1-f$, which
give the occupation of the final state of the collision \cite{Baym}. The local
density approximation is valid when $r_F p_F \gg \hbar$ where $r_F$ and $p_F$
are the sizes in position and momentum space of the Fermi gas. For a
temperature much lower than the Fermi temperature $T_F$ (where
$k_BT_F=(6N)^{1/3}\; \hbar
\omega$), this requires $N\gg1$.   Indeed, one gets in  this case
$r_F=(48N)^{1/6}a_{\rm HO}$
and $p_F=(48N)^{1/6}p_{\rm
HO}$, where $a_{\rm HO}=(\hbar/(m\omega))^{1/2}$ and
$p_{\rm HO}=(m\hbar\omega)^{1/2}$ are the spatial and momentum
extensions of the ground state harmonic oscillator.

We shall assume that the $w(\bf r,\bf p)$ is initially a distribution centered
in ${\bf r}_0,{\bf p}_0$, much narrower than the steady-state distribution of
(\ref{Bolt}): $w_{\rm ss} \propto \exp(-(p^2/2M + U(r))/(k_BT))$. As the
collisional relaxation proceeds the population of the narrow peak is
transferred to a broad distribution proportional to $w_{\rm ss}$. In the
following we focus on the initial stage of
 this relaxation phenomenon, namely the decay of the narrow peak, which
occurs at the rate $\Gamma_T({\bf r}_0,{\bf p}_0)$ deduced from (\ref{Bolt}):
\begin{equation}
 \Gamma_T({\bf r}_0,{\bf p}_0)= \frac{\sigma}{4\pi h^3}
\int d^3p_f\; d^2 \Omega\;
f(1-f')\;|{\bf v}_0-{\bf v}_f|
\label{Gamm}
\end{equation}
This rate could also be derived from the Fermi-Golden-Rule, within
the local density approximation.

We now consider three different classes of trajectories for P and discuss how
the damping of those trajectories is affected by the Fermi statistics of the
cloud. Most of the calculations assume equal mass for the P and the fermionic
atoms. Such a condition is approximately realized if P and the fermions are
isotopes of the same element ({\it i.e.} $^7$Li and $^6$Li, $^{39}$K and
$^{40}$K). For each trajectory class we calculate the multiple integral
(\ref{Gamm}) for an arbitrary temperature numerically and we derive scaling
laws for interesting limiting cases.

The first situation consists of P at rest in the trap center. For a Boltzmann gas with the same number of atoms, we
would expect $\Gamma^{\rm (Bol.)}_T(0,0)=n\sigma v_{\rm th}$, where $n$ is the spatial density at the center of the
trap for this gas and $v_{\rm th}$ is the most probable speed ($v_{\rm th}=(8k_BT/(\pi m))^{1/2}$). To put in evidence
the effects of the Fermi statistics, we plot in Fig.\ref{figure1} the rate $\Gamma_T(0,0)$ normalized by $\Gamma^{\rm
(Bol.)}_T(0,0)$. At high temperature ($T>T_F$) the deviations due to Fermi statistics are negligible. On the other
hand, for $T<T_F$, these deviations are spectacular and we find that $\Gamma_T/\Gamma^{\rm (Bol.)}_T \propto T^3$. This
power law dependence originates from two different phenomena both related to Fermi statistics. (i) For the Boltzmann
gas, $n \propto T^{-3/2}$ and $v\propto\sqrt{T}$, while in the fermionic gas the spatial density and the most probable
speed remain constant for vanishing $T$. This gives account for a factor $\propto T$. (ii) For $T \ll T_F$, P has a
finite collision probability only with fermions within an energy interval $\Delta E \sim k_B T$ (see
Fig.\ref{figure2})at the surface of the Fermi sphere, \cite{therm}. The solid angle $\Delta \Omega$ available for the
allowed fermionic final states is also proportional to $T$. Therefore the collisional rate is reduced by an additional
factor $\Delta E \; \Delta \Omega \propto T^2$.

This situation is well suited for determining the temperature of the Fermi gas
in the degenerate regime. Its absolute calibration depends on the exact mass
ratio between P and the fermions. Consequently in Fig.\ref{figure1} we also
plot $\Gamma_T(0,0)$ for the specific case of a cesium atom ($^{133}$Cs) as P
for a lithium gas ($^6$Li).

We now consider a second type of trajectory consisting of an oscillation of P
with an energy $E$, and a angular momentum $L$ which we set equal to zero (see
insert of Fig.\ref{figure3}). To calculate the rate at which P is ejected from
this trajectory by collisions, we suppose that the relaxation is slow with
respect to the period of an oscillation (collisionless regime). Since
$\Gamma_T({\bf r},{\bf p})$ is not constant over the trajectory, we define an
average collision rate:
\begin{equation}
{\gamma}_T(E,L=0)=\frac{\omega}{2\pi}\oint
\Gamma_T(r(t),p(t)) \; dt \quad .
\end{equation}
Depending on the energy of the oscillation and the temperature of the fermionic
gas, we find different regimes (see Fig.\ref{figure3}). As expected, for
$T>T_F$ the relaxation rate is the same as the one predicted for a Boltzmann
gas. It doesn't depend much on the energy of the excitation even for $E>k_BT$
(as long as we assume a constant $s$-wave elastic cross section). Indeed, when
$E$ grows the relative velocity between P and the cloud increases as $E^{1/2}$
while the fraction of the time P spends within the cloud decreases as
$E^{-1/2}$, leading to a constant average rate. For the degenerate case
$T<T_F$, three energy domains have to be considered for P. (i) For $E<k_BT<E_F$
we recover the rate $\Gamma_T(0,0)$ displayed in Fig.\ref{figure1}. (ii) For
$k_BT<E<E_F$, the rate varies as $E^2$. This can be easily understood at zero
temperature with $E \ll E_F$. In this case only collisions with fermionic
particles with energy close to $E_F$ contribute (first factor $E$), and the
solid angle available for the allowed final states brings an extra factor $E$.
(iii) For $k_BT<E_F<E$, all final states for the fermions after the collision
lie above the Fermi surface so that the inhibition due to statistics in no
longer effective. One recovers in this case a rate independent from $E$ as for
a Boltzmann gas.

The last type of trajectory considered in this paper consists of a circular
orbit of the test particle in the trapping potential ($L=E/\omega$). The
corresponding decay rate is plotted in Fig.\ref{figure4} as a function of the
mechanical energy of P. For a weakly degenerate gas we recover the same result
as for a Boltzmann gas. When one increases the energy of P, the damping rate is
constant up to $k_BT$ and then decreases, as P is orbiting outside the cloud.
In the degenerate case the damping rate presents a resonant behavior around
$E=E_F$. For $E\ll E_F$ this rate is decreased because of Pauli's exclusion
principle, while for $E\gg E_F$ it is small since P is outside the Fermi cloud.

We now address the preparation of P with arbitrary initial position and
momentum. The simplest idea is to exploit the difference in the ground state
hyperfine splitting between the various alkali atoms, or between two isotopes
of the same species. Consider for example the specific case of Lithium atoms.
If one starts with a mixture of $^6$Li and $^7$Li, one can perform
radiofrequency evaporation around a frequency $\nu_{\rm r.f.}\sim\;804$~MHz,
corresponding to the $^7$Li hyperfine splitting. One can prepare in this way an
ultracold sample of $^7$Li, playing the role of P, without eliminating any
fermionic atom $^6$Li, whose hyperfine splitting corresponds to $228$~MHz
\cite{sympathic}. The state obtained in this way corresponds to the first
situation considered in this paper. One can then prepare P on an arbitrary
trajectory by means of successive optical stimulated Raman transitions
\cite{Kasevich}. Due to the isotopic shift of the Li resonance line (10~GHz),
these transitions can be made isotopically selective. In a realistic case, we
can consider $10^8$ fermions in an isotropic magnetic trap with
$\omega/2\pi=100$~Hz ($T_F=4\;\mu$K, $r_F=0.17\;$mm). The density of the
fermionic gas is $5\; 10^{12}$~cm$^{-3}$, giving a mean field energy created by
the Fermi cloud on P equal to 10 nK \cite{symvanabeelen,houbiers}, which is
negligible with respect to $E_F$, as assumed in this paper.

To summarize we have shown in this paper that the collisional relaxation of a
probe particle imbedded in a Fermi atomic cloud gives a direct access to the
quantum degeneracy of this gas. In the degenerate regime P can be regarded as
an excitation ``frozen" by Pauli's exclusion principle. One can determine both
the temperature of the Fermionic cloud from the value of the relaxation rate
$\Gamma$ for $E \sim 0$, and the value of the Fermi energy exploiting the
resonant behavior of $\Gamma$ for $E\sim E_F$. For simplicity we have
considered here a non-interacting Fermi gas, but it is clear that this method
can be extended to study the effects of interactions onto the fermionic
excitation spectrum. In particular we plan to address the consequences of
Cooper pairing of the fermions \cite{Stoof1,stoof2,baranov,leggett} in a
subsequent paper.

I am very grateful to Jean Dalibard for stimulating
discussions and careful reading of the notes. I acknowledge useful
discussions with Micha Baranov, Yvan Castin, Marc-Oliver Mewes,
Christophe Salomon and
Florian Schreck.

This work was partially supported by CNRS, Coll\`{e}ge de France,
DRET, DRED and EC (TMR network ERB FMRX-CT96-0002).

\newpage

\begin{figure}
%\scalebox{0.6}{
%\includegraphics{figure1.eps}}
\caption{ Collisional rate $\Gamma_T({\bf r}=0,{\bf p}=0)$ of  a probe particle
(P) of mass $M$ at rest in the bottom of
a trap containing a fermionic gas of temperature $T$ (atomic mass $m$).
The temperature is plotted in units of the
Fermi temperature $T_F$, and $\Gamma_T$ in units of the rate
$\Gamma^{\rm Bol.}$ for a Boltzmann gas at the same
temperature and with the same number of atoms. Squares: $M=m$,
 circles: $M=(133/6)m$ (case of a $^6$Li gas probed by
$^{133}$Cs).} \label{figure1}
\end{figure}

\begin{figure}
%\begin{center}
%\scalebox{0.45}{
%\includegraphics{figure2.eps}}
%\end{center}
\caption{Collision between a fermion and P at rest. The fermion
has an initial  momentum equal to the Fermi momentum
$p_F$. $p'_P$ and $p'_f$ are the momenta of P and the fermion
after collision. The circumference with diameter $p_F$
passing through the centre of the Fermi sphere gives the possible
final states in the case of equal mass. The shell of
thickness $mk_BT/p_F$ represents the final states available to the
 Fermi particle.} \label{figure2}
\end{figure}

\begin{figure}
%\scalebox{0.5}{
%\includegraphics{figure3.eps}}
\caption{Damping rate $\gamma_T(E,L=0)$ for a linear oscillation
of P as a function of its excitation energy $E$. $E$
is expressed in units of the Fermi energy and $\gamma_T$ is
normalized by the rate $n\sigma v_{F}$, where $n$ is the
fermionic density at the bottom of the trap and
 $v_F$ the Fermi velocity. Circles: $T=T_F$, squares: $T=T_F/4$,
diamonds: $T=T_F/16$, triangles: $T=T_F/64$. The
inset shows a trajectory of P through the fermionic cloud. }
\label{figure3}
\end{figure}

\begin{figure}
%\scalebox{0.6}{
%\includegraphics{figure4.eps}}
\caption{Damping rate $\gamma_T(E,L=E/\omega)$ for a
circular orbit of the TP
as a function of its excitation energy $E$.
The normalizations and symbols are the same as in Fig.3.}
\label{figure4}
\end{figure}


\begin{thebibliography}{99}
\bibitem{therm}
These collisions transfer to P an energy $\sim k_B T$ thus ensuring its thermalization with the fermionic gas.
\end{thebibliography}
 
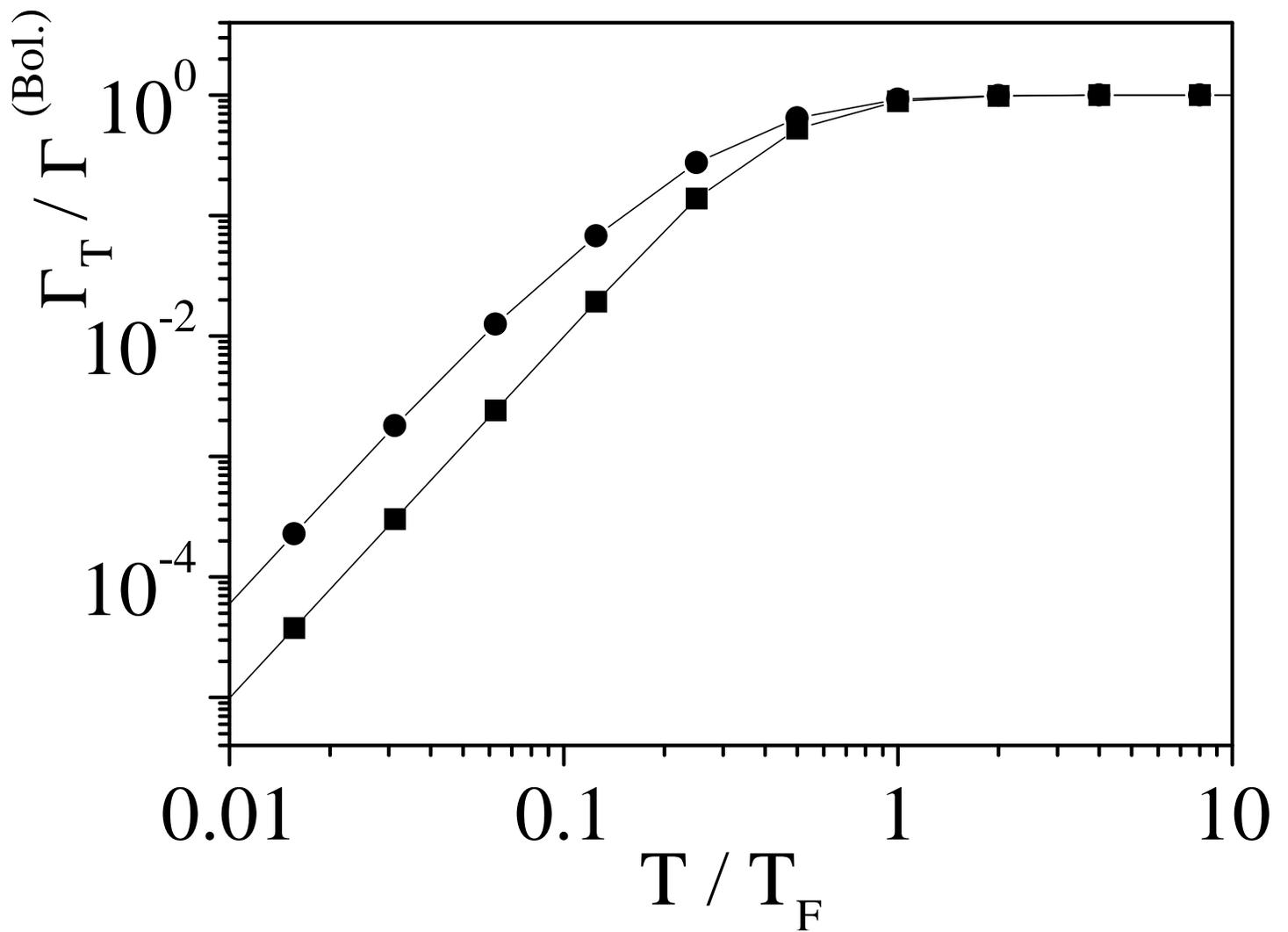
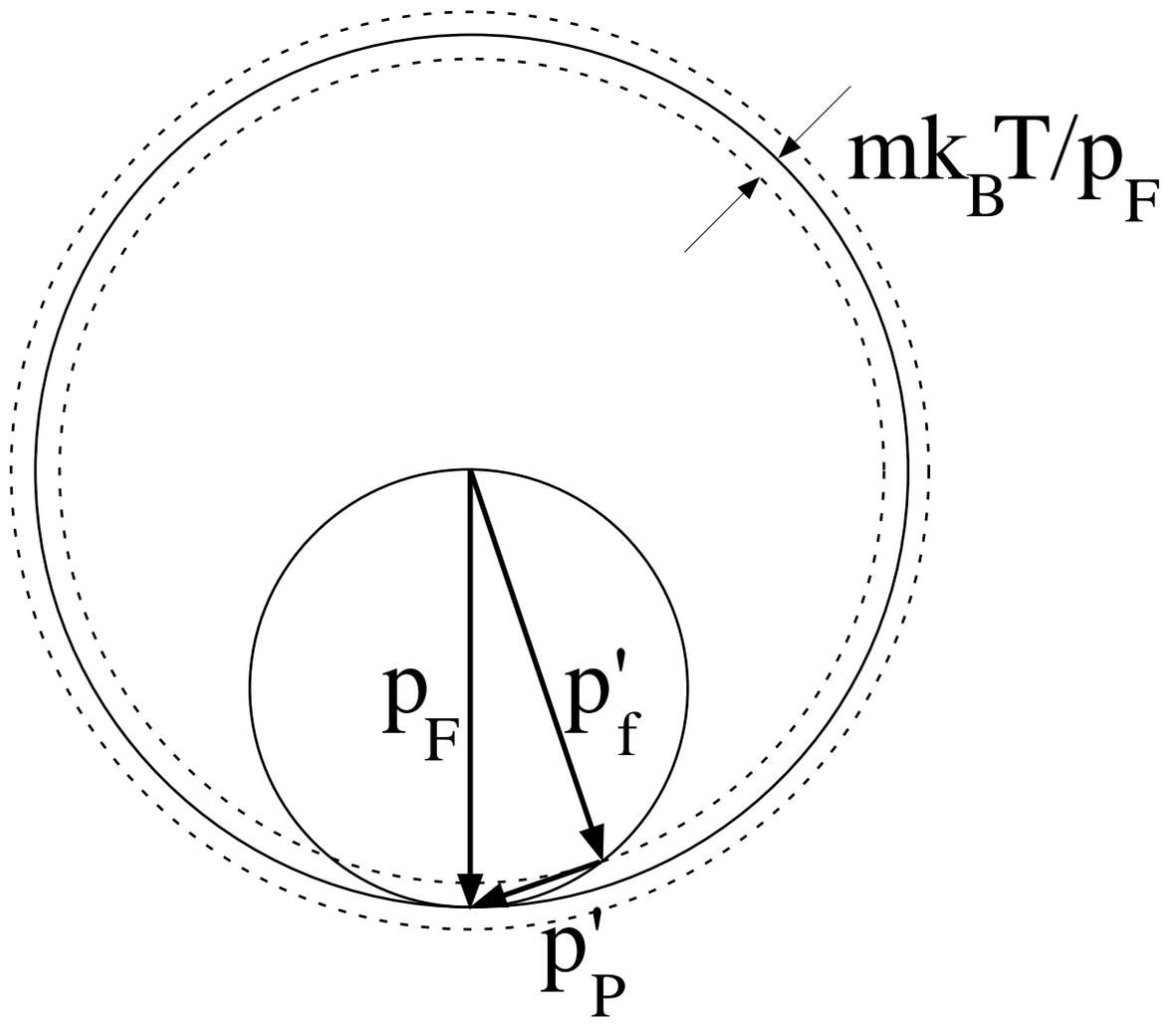
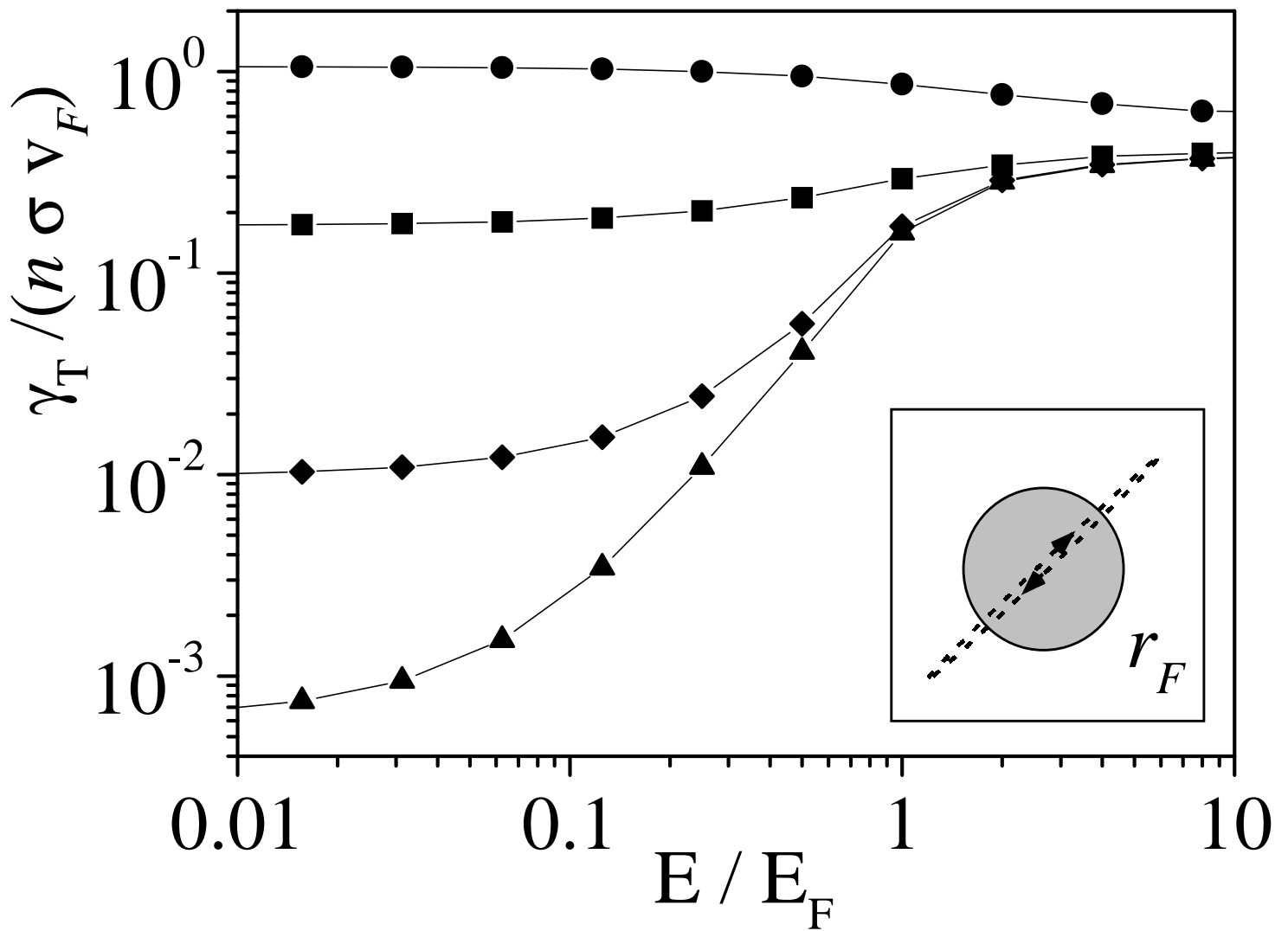
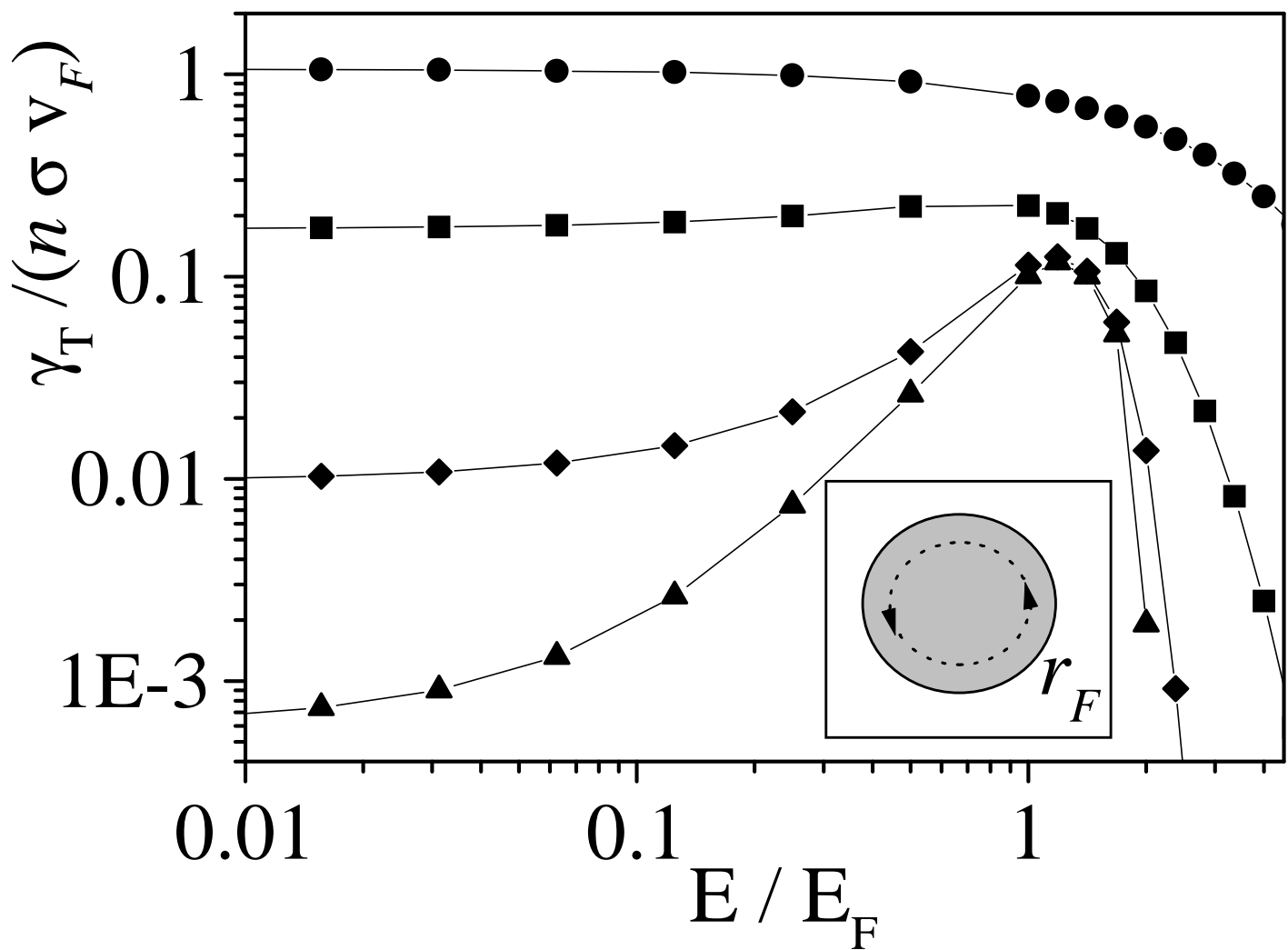
\end{document}